\documentclass[prl,aps,floats,twocolumn,showpacs,
preprintnumbers,showkeys]{revtex4-1}

\usepackage[latin1]{inputenc}                    
\usepackage{graphicx}                            
\usepackage{latexsym}                            
\usepackage{amsfonts}                            
\usepackage{amssymb}                             
\usepackage{amsmath}                             
\usepackage[mathscr]{eucal}                      
\usepackage{dcolumn}                             
\usepackage{theorem}                             
\newcommand{\bc}{\begin{center}}
\newcommand{\ec}{\end{center}}
\newcommand{\be}{\begin{equation}}
\newcommand{\ee}{\end{equation}}
\newcommand{\bea}{\begin{eqnarray}}
\newcommand{\eea}{\end{eqnarray}}
\newcommand{\non}{\nonumber}

\begin{document}

\preprint{
\vbox{
\hbox{ADP-10-22/T744}
}}

\title[Mass of the H-dibaryon]{Mass of the {\it H}-dibaryon}
\author{P.E.~Shanahan}
\author{A.W.~Thomas}
\author{R.D.~Young}
\affiliation{CSSM and CoEPP, School of Chemistry and Physics,
University of Adelaide,
  Adelaide SA 5005, Australia}

\begin{abstract}
Recent lattice QCD calculations have reported evidence for 
the existence of a bound state with strangeness $-2$ and baryon number 2 
at quark masses somewhat higher than the physical values. 
By developing a description of the 
dependence of this binding energy on the up, down and strange quark masses  
that allows a controlled chiral extrapolation,  
we explore the hypothesis that this state is to be 
identified with the $H$-dibaryon. Taking 
as input the recent results of the HAL and NPLQCD Collaborations,  
we show that the $H$-dibaryon is likely to be unbound 
by $13 \pm 14$~MeV at the physical point.
\end{abstract}



\maketitle


Our understanding of quantum chromodynamics has been 
challenged for decades by the apparent 
absence of multi-quark states. The outstanding candidate for such 
a state has been the $H$-dibaryon, ever since it was suggested that it should 
be very deeply bound with respect to the $\Lambda - \Lambda$ threshold~\cite{Jaffe:1976yi}.
Extensive experimental efforts to find this new 
particle~\cite{Yoon:2007aq,Takahashi:2001nm,Stotzer:1997vr,Iijima:1992pp} have led to the 
conclusion that it does not appear to be bound. However, the issue has been given new 
life in the past few months by reports from the HAL and NPLQCD Collaborations, whose 
lattice simulations find that the $H$-particle is indeed bound at quark masses somewhat 
above the physical range~\cite{Inoue:2010es,Beane:2010hg}. 

Early spectroscopic studies of the baryon and di-baryon spectrum within the bag 
model~\cite{Mulders:1982da} showed that the pion cloud contribution was  
phenomenologically very important in determining whether or not the $H$ was 
indeed bound. A preliminary analysis of some of the lattice data using a fit 
linear in $m_\pi^2$ suggests that it is bound at the physical quark mass, while 
an extrapolation linear in $m_\pi$, while also consistent with binding, does allow 
that the $H$ may be slightly unbound~\cite{Beane:2011xf}. We consider it important that 
the extrapolation to the physical quark masses should respect the constraints of chiral 
symmetry, such as ensuring the correct leading non-analytic behavior. 
Given the rather large 
range of quark masses over which current lattice simulations have been made, the 
technique which offers the best opportunity for a 
quantitative fit, while preserving the 
correct non-analytic behavior, is finite range 
regularization (FRR)~\cite{Leinweber:2003dg,Young:2002ib,Young:2002cj}. 

In this Letter we explore the possibility that the $H$ is a compact, multi-quark state that may 
be bound with respect to the $\Lambda-\Lambda$ threshold. We 
apply the FRR technique to describe the quark-mass dependence of both the octet-baryon masses and the binding 
energy of the $H$-dibaryon. With few lattice results available for the dibaryon binding energy,
it is essential to utilise the fit to the hyperon masses to determine the dependence on the non-singlet
combination of quark masses (i.e., $m_l-m_s$), which plays a critical role in describing the 
variation of the (non-singlet) $\Lambda$ mass. 
Extrapolated to the physical quark masses, 
we conclude that the $H$-dibaryon is 
most likely unbound, with its mass being $13 \pm 14$ MeV above the $\Lambda - \Lambda$ 
threshold.

{}Following the technique described in Ref.~\cite{Young:2009zb}, we fit the data for octet masses recently published by 
the PACS-CS Collaboration~\cite{Aoki:2008sm}, using an expansion about the SU(3) limit for  
the light and strange quark masses:
\be
M_B=M^{(0)}+\delta M_B^{(1)}+\delta M_B^{(3/2)}+ \ldots
\label{eq:1}
\ee
Here the leading term, $M^{(0)}$, denotes the degenerate mass of the baryon octet in 
the SU(3) chiral limit, and 
\be
\delta M_B^{(1)} = -C_{Bl}^{(1)} m_l - C_{Bs}^{(1)} m_s,
\label{eq:2}
\ee
with the coefficients given in Table~\ref{table1}, is the correction linear in the 
quark masses.
\begin{center}
\begin{table}
\begin{tabular}{c c c}
\hline
$B$ & $C_{Bl}^{(1)}$ & $C_{Bs}^{(1)}$ \\ \hline
$N$ & $2\alpha+2\beta+4\sigma$ & $2\sigma$ \\
$\Lambda$ & $\alpha + 2\beta + 4\sigma$ & $\alpha+2\sigma$ \\
$\Sigma$ & $\frac{5}{3}\alpha + \frac{2}{3}\beta + 4\sigma$ & $\frac{1}{3}\alpha+ \frac{4}{3}\beta + 2\sigma$ \\
$\Xi$ & $\frac{1}{3}\alpha+\frac{4}{3}\beta+4\sigma$ & $\frac{5}{3}\alpha+\frac{2}{3}\beta+2\sigma$ \\ \hline
\end{tabular}
\caption{Values for the terms linear in the non-strange quark mass, 
$m_l \rightarrow m_\pi^2/2$, and the strange quark mass, $m_s \rightarrow (m_K^2 - m_\pi^2/2)$,
expressed in terms of the leading quark-mass insertion parameters $\alpha , \beta$ and $\sigma$.}
\label{table1}
\end{table}
\end{center}

At next order, after the linear mass insertions, one finds quantum corrections associated with the 
chiral loops involving the pseudo-Goldstone
bosons, $\phi \equiv \pi$, $K$, $\eta$. While our formal assessment of the order of a given 
diagram treats the intermediate octet and decuplet baryons as degenerate, in order to more 
accurately represent the branch structure near $m_\phi \sim \delta$, we retain the octet-decuplet 
mass difference ($\delta$) in the numerical evaluations. 
These loops take the form:
\bea
\delta M_B^{(3/2)} = -\frac{1}{16\pi f^2} \sum_\phi &\left[\chi_{B\phi}I_R(m_\phi,0,\Lambda)\right.\non \\
& \left. +\chi_{T\phi}I_R(m_\phi,\delta,\Lambda)\right] \, ,
\label{eq:3}
\eea
where the coefficients $\chi_{B\phi}$, $\chi_{T\phi}$, 
taken from Ref.~\cite{WalkerLoud:2004hf}, are given in Table~\ref{table2}.
\begin{center}
\begin{table*}[tbh]
\begin{tabular}{c | c c c | c c c}
\hline
 & & $\chi_{B\phi}$ & & & $\chi_{T\phi}$ & \\
 & $\pi$ & $K$ & $\eta$ & $\pi$ & $K$ & $\eta$ \\ \hline
$N$ & $\frac{3}{2}(D+F)^2$ & $\frac{1}{3}(5D^2-6DF+9F^2)$ & $\frac{1}{6}(D-3F)^2$ & $\frac{4}{3}C^2$ & $\frac{1}{3}C^2 $ & 0 \\
$\Lambda$ & $2D^2$ & $\frac{2}{3}(D^2+9F^2)$ & $\frac{2}{3}D^2$ & $C^2$ & $\frac{2}{3}C^2$ & 0 \\
$\Sigma$ & $\frac{2}{3}(D^2+6F^2)$ & $2(D^2+F^2)$ & $\frac{2}{3}D^2$ & $\frac{2}{9}C^2$ & $\frac{10}{9}C^2$ & $\frac{1}{3}C^2$ \\
$\Xi$ & $\frac{3}{2}(D-F)^2$ & $\frac{1}{3}(5D^2+6DF+9F^2)$ & $\frac{1}{6}(D+3F)^2$ & $\frac{1}{3}C^2$ & $C^2$ & $\frac{1}{3}C^2$ \\ \hline
\end{tabular}
\caption{Chiral SU(3) coefficients for the octet baryons to octet ($B$) and decuplet ($T$) baryons 
through the pseudoscalar octet meson $\phi$.}
\label{table2}
\end{table*}
\end{center}
The meson loops involve the integrals:
\be
I_R=\frac{2}{\pi}\int dk \frac{k^4}{\sqrt{k^2+m_{\phi}^2}(\delta + \sqrt{k^2+m_{\phi}^2})}u^2(k)-b_0-b_2m_{\phi}^2
\label{eq:4}
\ee
where the subtraction constants, $b_{0,2}$, are defined so that the parameters $M^{(0)}, C_{Bl}^{(1)}$ and 
$C_{Bs}^{(1)}$ are renormalized
(explicit expressions may be found in Ref.~\cite{Leinweber:2003dg}, or can be readily evaluated numerically
by Taylor expanding the integrand in $m_{\phi}^2$).
The loop contribution parameters are taken to be $D+F=g_A=$~1.27, $F=\frac{2}{3}D$, 
$C=-2D$, $f=0.0871$~GeV and $\delta=0.292$~GeV. 
Within the framework of FRR, we introduce a mass scale, $\Lambda$, through  
a dipole regulator, $u(k)=(\frac{\Lambda^2}{\Lambda^2+k^2})^2$. 
$\Lambda$ is related to the 
scale (typically $\sim \Lambda/3$ for a dipole) beyond which a formal expansion in 
powers of the Goldstone boson masses breaks down~\cite{Lepage:1997cs,Young:2004tb}. 
However, rather than growing uncontrollably as some power of $m_\phi^2$, 
in this regime the 
Goldstone loops are actually suppressed, decreasing as powers of 
$\Lambda/m_\phi$~\cite{Thomas:2002sj,Stuckey:1996qr,Donoghue:1998bs,Leinweber:1998ej}.
In practice, we choose this mass parameter by fitting the lattice data itself.
Extensive studies of the FRR technique have established that the extrapolation is 
independent of the functional form chosen for 
the regulator~\cite{Leinweber:2003dg} -- essentially 
because of the rapid decrease of the loop contributions which we just explained.

In order to describe the mass of the $H$, treated as a compact, multi-quark state 
(rather than a loosely bound molecular state), we note that because it is an SU(3) singlet, 
at the equivalent order of the quark mass expansion its mass can be expressed as:
\be
M_H=M^{(0)}_H-\sigma_H(\frac{m_\pi^2}{2}+m_K^2)+\delta M^{(3/2)}_H \, ,
\label{eq:mH}
\ee
with
\bea
\delta M_H^{(3/2)} =& \! \! \! \! \! \! \! \! \! \! \! \! \! \!-C_H(2D^2+C^2) \left[I_R(m_\pi,\delta H,\Lambda)\right.\non \\
 &\left.+\frac{4}{3}I_R(m_K,\delta H,\Lambda)+\frac{1}{3}I_R(m_\eta,\delta H,\Lambda)\right] \, .
\label{eq:3.2}
\eea

As indicated in Eq.~(\ref{eq:mH}), because it is a flavour-singlet, at leading order in the quark masses $M_H$ depends only on the sum
of the quark masses. We therefore set:
\bea
B_H & = & 2M_\Lambda-M_H \non \\
& = & (2M^{(0)} - M^{(0)}_H)-(4\sigma_\Lambda-\sigma_H)(\frac{m_\pi^2}{2}+m_K^2) \non \\
&& -2\alpha m_K^2-2\beta m_\pi^2 +2\delta M^{(3/2)}_\Lambda - \delta M^{(3/2)}_H \non \\
& = & B_0 -\sigma_B(\frac{m_\pi^2}{2}+m_K^2)- 2\alpha m_K^2 -2\beta m_\pi^2  \non \\
&& +2 \delta M^{(3/2)}_\Lambda - \delta M^{(3/2)}_H \, ,
\label{eq:5}
\eea
where $B_0$ and $\sigma_B$ are parameters determined by the fit to the lattice
data for $B_H$~\cite{Inoue:2010es,Beane:2010hg}. Of course, $\alpha$ and $\beta$ are determined by the fit to
the baryon octet described above. For simplicity, 
we keep the regulator mass for the octet and 
the $H$ the same, while varying the chiral coefficient for 
the $H$, $C_H$, to fit the lattice data.

\begin{figure}[tb]
\bc
\includegraphics[width=0.49\textwidth]{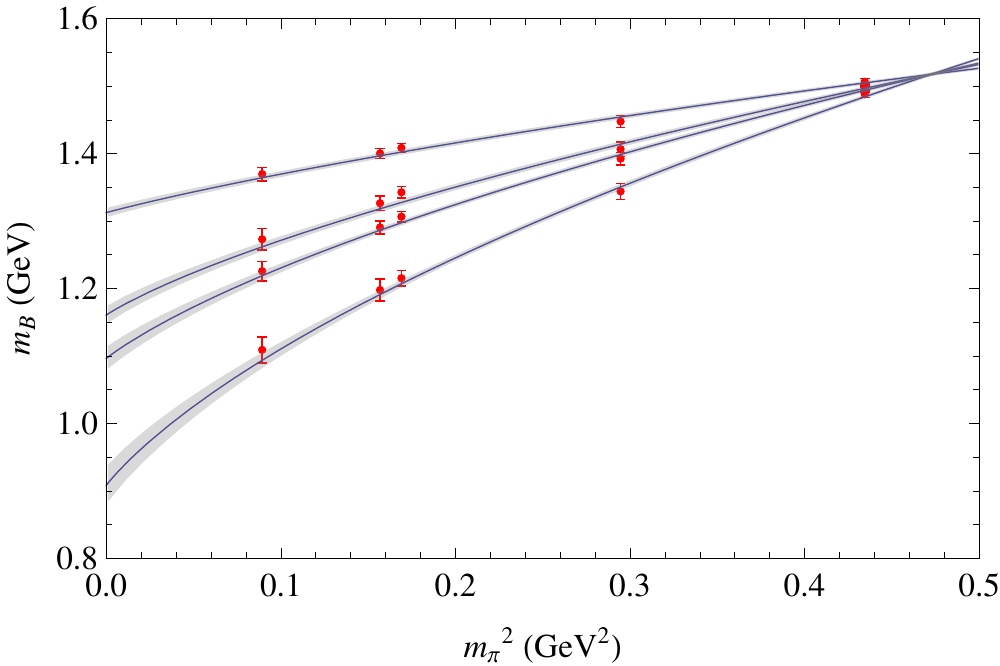}
\caption{Fit to the octet data of PACS-CS~\cite{Aoki:2008sm} using Eq.~(\ref{eq:1}). Note that 
we have fit the data after applying finite volume corrections and we have also used our fit 
to correct the lattice data for the strange quark mass, which was somewhat larger than the 
physical value.}
\label{fig:1}
\ec
\end{figure}

In order to fully maintain the correlations between 
the errors associated with all of the 
{}fitting parameters, we carried out a simultaneous analysis 
by minimizing $\chi^2$ for the 
{}fit to both the masses of the nucleon octet and 
to the difference in mass of the $H$ 
and two $\Lambda$ hyperons. Of course, 
the parameters $M^{(0)}$, $\alpha$, $\beta$, $\sigma$, as well as the regulator mass 
$\Lambda$, were primarily determined by the fit 
to the PACS-CS data, which is shown in Fig.~\ref{fig:1}, with the corresponding 
parameters given in Table~\ref{table3}. We note that, as explained 
in Ref.~\cite{Young:2009zb}, 
the octet data was corrected for small, model independent finite volume 
effects before fitting. In the figure we show the lattice 
data after applying both the finite
volume correction and a correction (based upon our fit parameters) arising because the 
PACS-CS simulations used values of the strange quark mass that 
were somewhat larger than the 
empirical values. The $\chi^2$ per degree of freedom for the octet data  
was $0.49$ (7.3 divided by $20-5 \equiv 15$). This is lower than unity as, 
without access to the original data, we 
cannot incorporate the effect of correlations between the lattice data. 
Nevertheless, the fit is clearly very satisfactory over the entire range of quark masses 
explored in the simulations 
and should provide an excellent basis for the study of the possible binding of the $H$-dibaryon.
Indeed, the masses of the $N, \, \Lambda, \, \Sigma$ and $\Xi$ baryons at the physical point  
are $(0.959 \pm 0.023 \, , 1.129 \pm 0.014 \, , 1.188 \pm 0.011 \, , 1.325 \pm 0.006)$ GeV, 
where all the errors include the correlated uncertainties of all the fit parameters, 
including the regulator mass, $\Lambda$. For comparison we note that the physical octet 
masses are $(0.939 \, , 1.116 \, , 1.193 \, , 1.318)$ GeV.
\begin{center}
\begin{table*}[tbh]
\begin{tabular}{c | c c c c c c c c}
\hline
 & $\Lambda (GeV)$ & $M_0$~(GeV) & $\alpha$ (GeV$^{-1}$) & $\beta$ (GeV$^{-1}$) & $\sigma$ (GeV$^{-1}$)& $B_0$~(GeV) & $\sigma_B$ (GeV$^{-1}$)& $C_H$~(GeV$^{-2}$) \\ \hline
best fit value & 1.02 & 0.861 & -1.71 & -1.20 & -0.51 & 0.019 & -2.36 & 5.65  \\
error & 0.06 & 0.037 & 0.12 & 0.10 & 0.05 & 0.004 & 0.20 & 0.09  \\ \hline
\end{tabular}
\caption{Values of the fit parameters for the octet and $H$-dibaryon 
data corresponding to the 
fits shown in Figs.~1 and 2.}
\label{table3}
\end{table*}
\end{center}

With respect to the $H$-dibaryon we have retained only 
the data from the HAL Collaboration~\cite{Inoue:2010es} which was 
generated on the largest lattice volume, namely $3.87$~fm. These data points correspond to large 
(degenerate) pseudoscalar masses, $1.015$, $0.837$ and $0.673$~GeV, for which the finite volume 
corrections are expected to be very small. Accordingly we used the reported values 
without applying any finite volume correction. 
In the case of the NPLQCD Collaboration~\cite{Beane:2010hg}, where the calculation was 
performed at $m_\pi = 389$ MeV and $m_K = 553$ MeV, we include 
in the fit only the value for the binding of the $H$ determined after 
their extrapolation to infinite volume.

We chose the mass splitting between the $H$-dibaryon and the other dibaryon states appearing 
in its chiral loop corrections to be the same as the octet-decuplet mass splitting used earlier,  
namely $\delta H = \delta = 0.292$~GeV. This is compatible with the 
estimates of Aerts {\it et al.}~\cite{Aerts:1977rw} calculated within 
the MIT bag model, as well as with the experimental absence of other nearby states.
The sensitivity of our fit to $\delta H$ is quite small, with an increase (decrease) 
of $\delta$ by 100 MeV increasing (decreasing) the mass difference $2 M_\Lambda - M_H$ by only 4 MeV.
This small shift is combined in quadrature with the error found from our chiral fit to yield 
the final, quoted error in the binding of the $H$.

\begin{figure}[tb]
\bc
\includegraphics[width=0.49\textwidth]{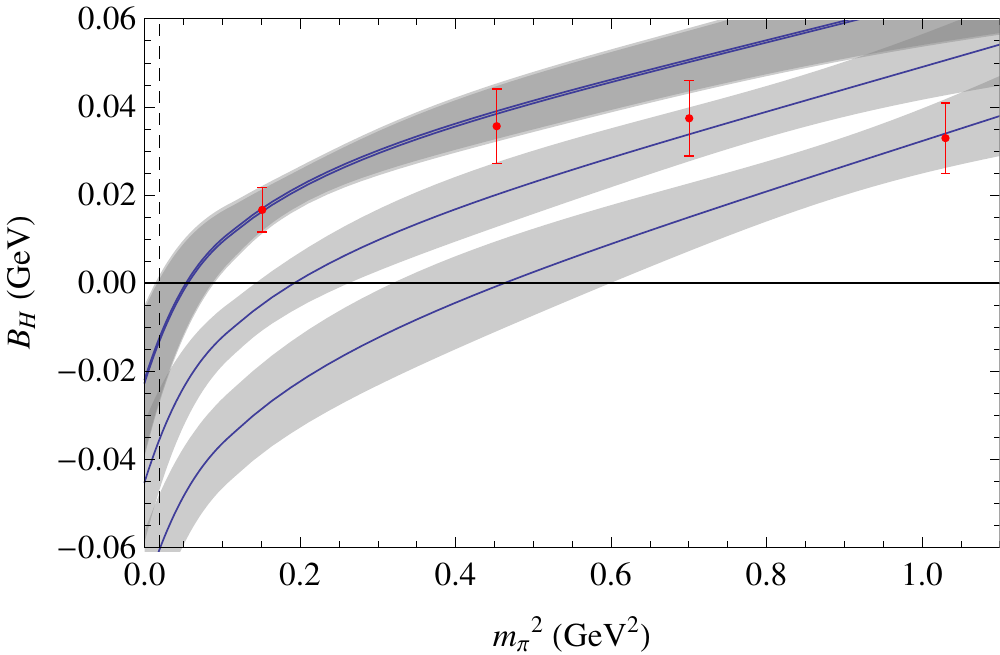}
\caption{Binding energy of the $H$-dibaryon versus pion mass squared, 
resulting from our 
chiral fit, for several values of the 
kaon mass at which the simulations of Refs.~\cite{Inoue:2010es} 
and \cite{Beane:2010hg} 
were carried out. 
}
\label{fig:2}
\ec
\end{figure}

The best fit parameters describing the binding energy of the $H$-dibaryon are 
also given in Table~\ref{table3}  
and the actual fit is shown in Fig.~\ref{fig:2}. As explained above, 
the data shown in Fig.~\ref{fig:2} are from NPLQCD 
(lowest mass point) and HAL (three largest mass points). In each case the curve nearest the data point 
illustrates the extrapolation as a function of the light quark mass implied by our fit at the value of the 
strange quark mass corresponding to that lattice data point. 
The errors shown are the result of combining in quadrature the statistical and systematic errors
quoted by the collaborations. The shaded error bands incorporate the effect of
correlations between the fit parameters, including the uncertainty on the regulator mass.
We note the remarkable result that the best fit value of 
the chiral coefficient for the $H$-dibaryon, $C_H$ 
(which for convenience is normalized with respect to the 
chiral coefficient for $\pi$ loops on the $\Lambda$ hyperon in Eq.~(\ref{eq:3.2})), 
is within 20\% of the theoretical value reported by Mulders and 
Thomas~\cite{Mulders:1982da}, who calculated it using SU(6) symmetry. 
If instead we retain the Mulders-Thomas coefficient, 
the $H$-dibaryon is unbound by $30 \pm 9$ MeV 
but the quality of the fit is significantly reduced, 
with a $\chi^2$ per degree of freedom of almost 2.

It is clear from Figs.~1 and 2 that both the octet data and the data 
on the binding energy 
of the $H$-dibaryon are very well described. 
The binding of the $H$ is reduced by a decrease in   
the masses of the $u$ and $d$ quarks and the $s$, with significant chiral curvature 
for $m_\pi$ below 0.4 GeV. 
It is important to note that our analysis does not include the effect of the 
coupling of the $H$-dibaryon to the open $\Lambda - \Lambda$ channel. 
Given where we find 
the multi-quark state, this might be either attractive or repulsive, depending on the 
range of momenta which dominate the continuum channel. It is also true that it would be 
extremely valuable to have new simulations at lower pion mass and large volume which 
would further constrain the extrapolation. 
Nevertheless, the conclusion of our study is quite clear. 
Even though the $H$ is bound at larger quark masses, we see in Fig.~\ref{fig:2} that 
chiral physics leads to a more rapid decrease of the mass of the $\Lambda$ as $m_\pi$ 
approaches its physical value than we find for the $H$ and as 
a result one must conclude that at the physical values of the quark
masses the $H$-dibaryon is most likely unbound. 
Our estimate, including the effect of
correlations between all the fit parameters, 
is that the $H$ is unbound by a mere $13 \pm 14$~MeV at
the physical point. That this is so close to the $\Lambda - \Lambda$ 
threshold will undoubtedly spur investigations into the consequences 
for doubly strange hypernuclei as well as the equation of state of dense matter.

%
\section*{Acknowledgements}
%
%
This work was supported by the University of Adelaide and the Australian
Research Council through grant FL0992247 (AWT) and DP110101265 (RDY).
%
%



\begin{thebibliography}{99}
%
\bibitem{Jaffe:1976yi}
R.~L.~Jaffe,
Phys.\ Rev.\ Lett.\  {\bf 38}, 195-198 (1977).
%
\bibitem{Yoon:2007aq}
C.~J.~Yoon, H.~Akikawa, K.~Aoki, Y.~Fukao, H.~Funahashi, M.~Hayata,
K.~Imai, K.~Miwa {\it et al.},
Phys.\ Rev.\  {\bf C75}, 022201 (2007).
%
\bibitem{Takahashi:2001nm}
H.~Takahashi, J.~K.~Ahn, H.~Akikawa, S.~Aoki, K.~Arai, S.~Y.~Bahk,
K.~M.~Baik, B.~Bassalleck {\it et al.},
Phys.\ Rev.\ Lett.\  {\bf 87}, 212502 (2001).
%
\bibitem{Stotzer:1997vr}
R.~W.~Stotzer {\it et al.} [ BNL E836 Collaboration ],
Phys.\ Rev.\ Lett.\  {\bf 78}, 3646-3649 (1997).
%
\bibitem{Iijima:1992pp}
T.~Iijima, H.~Funahashi, S.~Hirata, M.~Ieiri, I.~Imai, T.~Ishigami,
Y.~Itow, K.~Kikuchi {\it et al.},
Nucl.\ Phys.\  {\bf A546}, 588-606 (1992).
%
\bibitem{Inoue:2010es}
T.~Inoue {\it et al.} [HAL QCD Collaboration],
Phys.\ Rev.\ Lett.\  {\bf 106}, 162002 (2011).
%
\bibitem{Beane:2010hg}
S.~R.~Beane {\it et al.} [NPLQCD Collaboration],
Phys.\ Rev.\ Lett.\  {\bf 106}, 162001 (2011).
%
\bibitem{Mulders:1982da}
P.~J.~Mulders, A.~W.~Thomas,
J.\ Phys.\ G {\bf G9}, 1159 (1983).
%
\bibitem{Beane:2011xf}
S.~R.~Beane {\it et al.}, 
[arXiv:1103.2821 [hep-lat]].
%
\bibitem{Leinweber:2003dg}
D.~B.~Leinweber, A.~W.~Thomas, R.~D.~Young,
Phys.\ Rev.\ Lett.\  {\bf 92}, 242002 (2004).
%
\bibitem{Young:2002ib}
R.~D.~Young, D.~B.~Leinweber, A.~W.~Thomas,
Prog.\ Part.\ Nucl.\ Phys.\  {\bf 50}, 399-417 (2003).
%
\bibitem{Young:2002cj}
R.~D.~Young, D.~B.~Leinweber, A.~W.~Thomas, S.~V.~Wright,
Phys.\ Rev.\  {\bf D66}, 094507 (2002).
%
\bibitem{Aoki:2008sm}
S.~Aoki {\it et al.} [ PACS-CS Collaboration ],
Phys.\ Rev.\  {\bf D79}, 034503 (2009).
%
\bibitem{Young:2009zb}
R.~D.~Young, A.~W.~Thomas,
Phys.\ Rev.\  {\bf D81}, 014503 (2010).
%
\bibitem{WalkerLoud:2004hf}
A.~Walker-Loud,
Nucl.\ Phys.\  {\bf A747}, 476-507 (2005).
%
\bibitem{Thomas:2002sj}
A.~W.~Thomas,
Nucl.\ Phys.\ Proc.\ Suppl.\  {\bf 119}, 50-58 (2003).
%
\bibitem{Stuckey:1996qr}
R.~E.~Stuckey, M.~C.~Birse,
J.\ Phys.\ {\bf G23}, 29-40 (1997).
%
\bibitem{Donoghue:1998bs}
J.~F.~Donoghue, B.~R.~Holstein, B.~Borasoy,
Phys.\ Rev.\  {\bf D59}, 036002 (1999).
%
\bibitem{Leinweber:1998ej}
D.~B.~Leinweber, D.~-H.~Lu, A.~W.~Thomas,
Phys.\ Rev.\  {\bf D60}, 034014 (1999).
%
\bibitem{Lepage:1997cs}
G.~P.~Lepage,
[nucl-th/9706029].
%
\bibitem{Young:2004tb}
R.~D.~Young, D.~B.~Leinweber, A.~W.~Thomas,
Phys.\ Rev.\  {\bf D71}, 014001 (2005).
%
\bibitem{Aerts:1977rw}
A.~T.~M.~Aerts, P.~J.~G.~Mulders, J.~J.~de Swart,
Phys.\ Rev.\  {\bf D17}, 260 (1978).
%
\bibitem{Young:2009rj}
R.~D.~Young,
AIP Conf.\ Proc.\  {\bf 1182}, 905-908 (2009).
%
\end{thebibliography}
\end{document}